\documentclass[sigconf,authorversion,nonacm]{acmart}
\AtBeginDocument{%
  \providecommand\BibTeX{{%
    \normalfont B\kern-0.5em{\scshape i\kern-0.25em b}\kern-0.8em\TeX}}}

\setcopyright{acmlicensed}
\copyrightyear{2024}
\acmYear{2024}
\acmDOI{XXXXXXX.XXXXXXX}

\acmConference[ICPP '24]{International Conference on Parallel Processing}{August 12--15,
  2024}{Gotland, Sweden}
\acmBooktitle{Woodstock '18: ACM Symposium on Neural Gaze Detection,
 June 03--05, 2018, Woodstock, NY} 
\acmISBN{978-1-4503-XXXX-X/18/06}

\usepackage{graphicx}
\usepackage{algorithmicx}
\usepackage{textcomp}
\usepackage{xcolor}

\usepackage{multirow}
\usepackage{float}
\usepackage{listings}

\usepackage{siunitx}
\usepackage{booktabs}
\usepackage{hyperref}

\usepackage{array}

\usepackage{comment}

\usepackage{siunitx}
\usepackage{fp}
\usepackage{wrapfig}
\usepackage{subcaption}
\usepackage{sidecap}
\usepackage{graphicx}

\newcommand{\numtomillion}[1]{%
  \FPdiv{\result}{#1}{1000000}%
  \num[round-mode=places, round-precision=2, scientific-notation=false]{\result} M%
}

\newcommand{\numtothousand}[1]{%
  \FPdiv{\result}{#1}{1000}%
  \num[round-mode=places, round-precision=1, scientific-notation=false]{\result} K%
}

\usepackage{tikz}

\newcommand{\poolball}[1]{%
  \begin{tikzpicture}[baseline=(current bounding box.south)]
    \filldraw[fill=black, draw=black] (0,0) circle (0.18cm);
    \filldraw[fill=white, draw=black] (0,0) circle (0.14cm);
    
    \node at (0,0) {\scriptsize\textbf{#1}};
  \end{tikzpicture}%
}
\newcommand{\poolballtiny}[1]{%
  \begin{tikzpicture}[baseline=(current bounding box.south)]
    \filldraw[fill=black, draw=black] (0,0) circle (0.19cm);
    \filldraw[fill=white, draw=black] (0,0) circle (0.14cm);
    
    \node at (0,0) {\tiny\textbf{#1}};
  \end{tikzpicture}%
}

\NewDocumentCommand{\todo}{m}{%
    \textcolor{red}{(TODO: #1)}%
}

\NewDocumentCommand{\langkeyword}{m}{%
    \texttt{\textcolor{violet!80!black}{#1}}%
}

\NewDocumentCommand{\langoperator}{m}{%
    \texttt{\textcolor{orange!60!black}{#1}}%
}

\begin{document}

%\title{Structures and Techniques for Processing Highly Skewed Graphs on Message-Driven Systems}
%\title{Co-Designing A Message-Driven System for Processing Highly Skewed Graphs}
%\title{Programming Model and Runtime Support for Processing Highly Skewed Graphs on Message-Driven Systems}
\title{Rhizomes and Diffusions for Processing Highly Skewed Graphs on Fine-Grain Message-Driven Systems}
\author{Bibrak Qamar Chandio}
\email{bchandio@iu.edu}
\orcid{0009-0000-8228-6273}
\affiliation{%
  \institution{\textit{Department of Intelligent Systems Engineering}\\ Indiana University Bloomington}
  %\streetaddress{P.O. Box 1212}
  %\city{Bloomington}
  \state{Indiana}
  \country{USA}
  %\postcode{47408}
}

\author{Prateek Srivastava}
\email{pratsriv@iu.edu}
\affiliation{%
  \institution{\textit{Department of Intelligent Systems Engineering}\\ Indiana University Bloomington}
  %\streetaddress{P.O. Box 1212}
  %\city{Bloomington}
  \state{Indiana}
  \country{USA}
  %\postcode{47408}
}

\author{Maciej Brodowicz}
\email{mbrodowi@iu.edu}
\affiliation{%
  \institution{\textit{Department of Intelligent Systems Engineering}\\ Indiana University Bloomington}
  %\streetaddress{P.O. Box 1212}
  %\city{Bloomington}
  \state{Indiana}
  \country{USA}
  %\postcode{47408}
}

\author{Martin Swany}
\email{mswany@iu.edu}
\affiliation{%
  \institution{\textit{Department of Intelligent Systems Engineering}\\ Indiana University Bloomington}
  %\streetaddress{P.O. Box 1212}
  %\city{Bloomington}
  \state{Indiana}
  \country{USA}
  %\postcode{47408}
}

\author{Thomas Sterling}
\email{tron@iu.edu}
\affiliation{%
  \institution{\textit{Department of Intelligent Systems Engineering}\\ Indiana University Bloomington}
  %\streetaddress{P.O. Box 1212}
  %\city{Bloomington}
  \state{Indiana}
  \country{USA}
  %\postcode{47408}
}

\renewcommand{\shortauthors}{Bibrak et al.}

\begin{abstract}
The paper provides a unified co-design of 1) a programming and execution model that allows spawning tasks from within the vertex data at runtime, 2) language constructs for \textit{actions} that send work to where the data resides, combining parallel expressiveness of local control objects (LCOs) to implement asynchronous graph processing primitives, 3) and an innovative vertex-centric data-structure, using the concept of Rhizomes, that parallelizes both the out and in-degree load of vertex objects across many cores and yet provides a single programming abstraction to the vertex objects. The data structure hierarchically parallelizes the out-degree load of vertices and the in-degree load laterally. The rhizomes internally communicate and remain consistent, using event-driven synchronization mechanisms, to provide a unified and correct view of the vertex.

Simulated experimental results show performance gains for BFS, SSSP, and Page Rank on large chip sizes for the tested input graph datasets containing highly skewed degree distribution. The improvements come from the ability to express and create fine-grain dynamic computing task in the form of \textit{actions}, language constructs that aid the compiler to generate code that the runtime system uses to optimally schedule tasks, and the data structure that shares both in and out-degree compute workload among memory-processing elements.
\end{abstract}




%\received{29 April 2024}
%\received[revised]{X Month 2024}
%\received[accepted]{X Month 2024}

\maketitle

\lstdefinelanguage{Racket}{
    morekeywords={struct, begin, predicate, propagate, diffuse, define, work, if, cond, let, let*, for-each, set!, cons, Integer, Float, Vector, Pointer, null},
    keywordstyle={\color{violet!80!black}},
    sensitive=true,
    morecomment=[l]{;},
    morestring=[b]",
    alsoletter={<, >, !,-},
    % literate={<}{{$<$}}1 {>}{{$>$}}1 {!}{{!}}1 {lambda}{{$\lambda$}}1,
    % literate={lambda}{{$\lambda$}}1,
    morekeywords=[2]{SSSP-Action, SSSP-Diffuse, edge-address, edge-weight, vertex-edges, vertex-level, set-vertex-level!, list, inform-neighbors, bfs-action, BFSDiffuse,  set-future!, allocate, insert-edge-action, insert-edge, vertex-has-room, vertex-ghost, vertex-score, set-vertex-score!, future-pending, enqueue-future!, future-pending!, enqueue!, rhizome-collapse, op, bcast, vertex-iteration-score, vertex-msg-count, set-vertex-msg-count!, inform-score-to-neighbors, vertex-out-degree, vertex-in-degree},
    keywordstyle=[2]{\color{orange!60!black}},
    morekeywords=[3]{mutable, rhizome-shared},
    keywordstyle=[3]{\color{blue}},
    literate={lambda}{{{\color{violet!80!black}$\lambda$}}}1 
              {call/cc}{{{\color{violet!80!black}call/cc}}}6 
              {eq?}{{{\color{orange!60!black}eq?}}}3 
              {?}{{{\color{orange!60!black}?}}}1 
              {'}{{{\color{red!75!black}\small\textbf{\textquotesingle}}}}1 
              {==}{{{\color{orange!60!black}==}}}2 
              {+}{{{\color{orange!60!black}+}}}1 
              %{-}{{{\color{orange!60!black}-}}}1 
              {/}{{{\color{orange!60!black}/}}}1 
              {>}{{{\color{orange!60!black}>}}}1 
              {null?}{{{\color{orange!60!black}null?}}}5 
              {\#}{{{\color{blue}\#}}}1 
              {:}{{{\color{blue}:}}}1,
}

\section{Introduction}
Fine granularity of operations, large amount of latent parallelism, and irregular memory access are the properties of interest that have led to the work discussed in this paper. These properties are manifested in graph processing with the edges representing large amount of very fine-grain parallelism having no general locality in terms of where they might point, rendering popular techniques not as useful. These techniques include: anticipating and staging of data guided by the principle of spatial locality, bulk synchronous models of task expression and synchronization that impose or assume a coarser-granularity of operations, sending data to where the work needs to be performed, and static expression of parallelism rather than dynamic discovery at runtime from the graph data itself.

This paper focuses on alternate approaches such as asynchronous message-driven computations that have the potential to expose the inherent latent parallelism of the graph structure at runtime by send work to where the vertex resides. Such a message-driven system comes with its own challenges of a programming model and language that is easy to write vertex-centric programs, along with new data structures and runtime support for dynamic task creation, load balancing and parallelization of the underlying graph data.

In the above backdrop, we focus on and provide the following contributions:
\begin{itemize}
  \item A programming and execution model that allows spawning tasks from within the vertex data at runtime.
  \item An innovative vertex-centric data-structure, using the concept of Rhizomes, that parallelizes both the out-degree and in-degree load of vertex objects across many cores and yet provides a single programming abstraction to the vertex objects.
  \item Language constructs for \textit{actions} that send work to where the data resides, combining parallel expressiveness of Local Control Objects (LCOs) to implement asynchronous graph processing primitives.
  \item Language and compiler aided runtime system support for task creation and execution that keep compute resources occupied with work when messages block on network due to congestion.
\end{itemize}

\section{Message-Driven Systems}

We target message-driven near-memory systems that are generally categorized as being partitioned global address space (PGAS) many-cores tessellated in a network-on-chip (NOC), usually a mesh. Each core has a small amount of low-latency memory (usually SRAM), along with basic computational capability \cite{beyondstatic2023}\cite{Dalorex2023}\cite{celerity2017}\cite{GigaRisc2019}\cite{CCA2016}\cite{CCA2018}.

\begin{figure}
  \centering
  \includegraphics[width=1\linewidth]{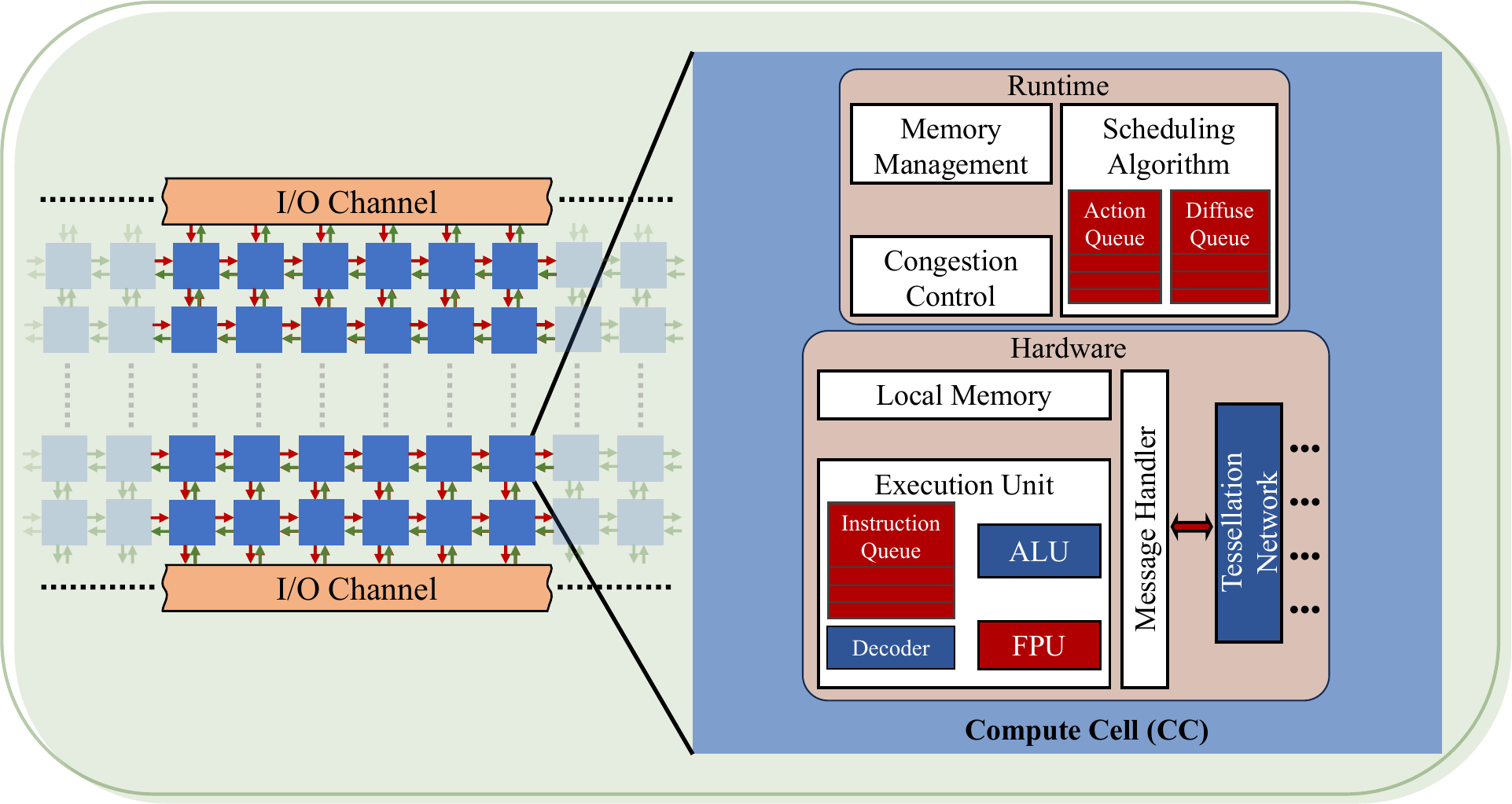}
  \captionsetup{skip=2pt} % Reduce the space between figure and caption
  \caption{An AM-CCA chip with many processing elements.}
  \label{fig:cca-chip-memory-controller}
\end{figure}

In particular, we use the Continuum Computing Architecture (CCA)\cite{CCA2016}\cite{CCA2018} as a design platform upon which we explore our ideas of data-structure, programming model, and runtime system. CCA is organized as an interconnection of homogeneous global memory Compute Cells (CCs), which are capable of; 1) data storage; 2) data manipulation; and 3) data transmission to adjacent CCs. Figure~\ref{fig:cca-chip-memory-controller} shows the architecture diagram of a CCA Chip, where an arrangement of CCs tessellated for tight coupling, and interconnected work together to provide logical unification, data storage capacity, and parallelism. Each CC contains interconnect links that interface its message handler, a limited capacity of local memory, and other resources for instruction execution. There is also a light-weight runtime system that we use to implement our ideas of memory management, congestion control, and programming-model-aided dynamic task creation and scheduling. We call our design the Active-Memory CCA (AM-CCA).

Although the scope of this paper is restricted to a single node, the underlying programming model is message-driven, and the memory model is PGAS. Therefore, we believe that the ideas discussed can be applied to multi-node distributed systems of many-cores, thereby exporting a single system programming image. This is in contrast to the more popular heterogeneous models, such as MPI+X, that have to patch together multiple programming models.

%Like HammerBlade \cite{beyondstatic2023}, the messages can arrive out-of-order creating synchronizing 

\section{Vertex-Centric Data Structure}
In its simplest form a vertex is composed of its local data, application-centric data, and edge-list containing out-edges. These out-edges are the pointers to addresses of other vertices in the graph and may contain any associated edge meta-data such as weight.

\begin{figure*}
  \centering
  \begin{subfigure}{.19\linewidth}
    \centering
    \includegraphics[width=\linewidth]{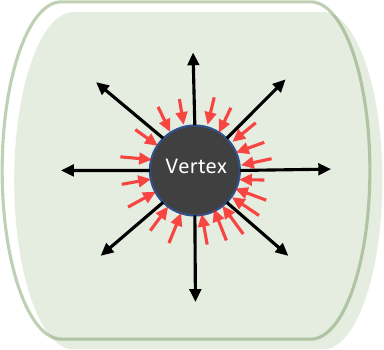}
    \caption{\textbf{V}ertex \textbf{O}bject (VO).}
    \label{fig:datastructure-vertex}
  \end{subfigure}
  \hfill
  \begin{subfigure}{.32\linewidth}
    \centering
    \includegraphics[width=\linewidth]{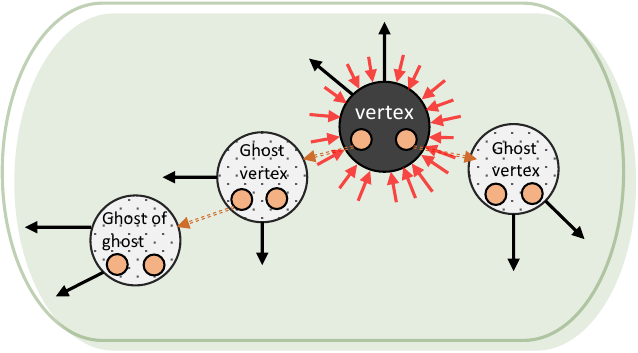}
    \caption{\textbf{R}ecursively \textbf{P}arallel \textbf{V}ertex \textbf{O}bject (RPVO).}
    \label{fig:datastructure-RPVO}
  \end{subfigure}
  \hfill
  \begin{subfigure}{.47\linewidth}
    \centering
    \includegraphics[width=\linewidth]{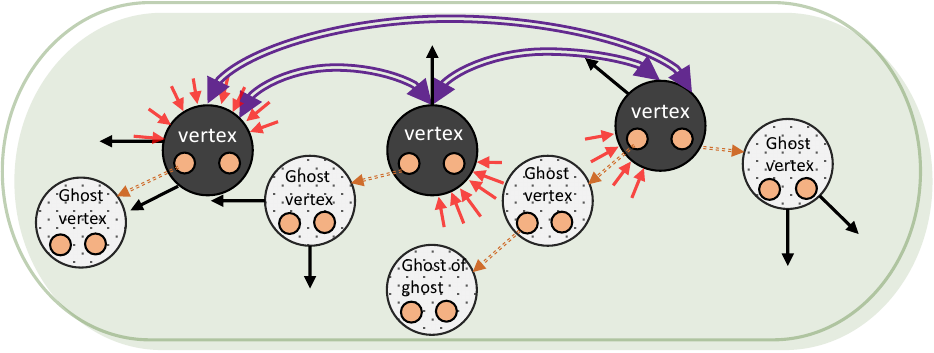}
    \caption{\textbf{Rh}izomatic \textbf{R}ecursively \textbf{P}arallel \textbf{V}ertex \textbf{O}bject (Rh-RPVO).}
    \label{fig:datastructure-rhizome}
  \end{subfigure}
  \hfill
  \begin{subfigure}{0.7\linewidth}
    \centering
    \includegraphics[width=\linewidth]{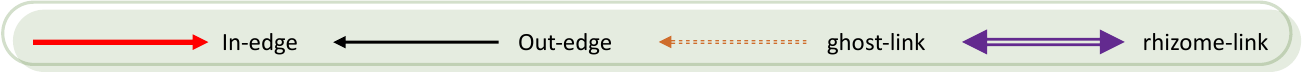}
    \label{fig:datastructure-rhizome-legend}
  \end{subfigure}
  \captionsetup{skip=0pt} % Reduce the space between figure and caption
  \caption{Vertex-Centric Data Structures: (a) a simple vertex having all its out-edges stored in a list and being pointed by all its in-edges, (b) the same vertex but with only its out-edges partitioned hierarchically, (c) the same vertex with not only its out-edges partitioned hierarchically but also its in-edges partitioned rhizomatically.}
  \label{fig:vertex-datastructures}
\end{figure*}

%as shown in Listing \ref{lst:data-structure}. 

\begin{comment}
%\setlength{\intextsep}{0pt} % Adjust vertical spacing between text and floats
%\setlength{\columnsep}{13pt} % Adjust horizontal spacing between columns
% Using wrapfigure to wrap text around the code listing
\begin{wrapfigure}[10]{l}{0.3\textwidth}
\begin{lstlisting}[language=Racket,caption=Vertex-Centric Data Structure,label=lst:data-structure,xleftmargin=0.3cm]
; vertex type for BFS
(struct vertex 
  ([id    : Integer] 
   [level : Integer]
   [edges : (Vector edge)]))
\end{lstlisting}
\end{wrapfigure}
\end{comment}

A graph composed of such vertices must be partitioned across cores (tiles), in manycore systems, or nodes in a distributed system. This is not just due to the need to expose and exploit parallelism but also to the sheer necessity caused by limited memory capacity at a given locality. In this regard, the problem is twofold: 1) partitioning and managing out-degree load, and 2) partitioning and managing in-degree load.

Recent works, such as Aspen \cite{aspen2019}, Terrace \cite{Terrace2021}, and GraphVine \cite{graphvine2023} address the former problem. In particular, designing a vertex data structure, for dynamic graphs, that stores out-edges of a large vertex in a tree-like structure. The central idea is to hierarchically partition any number of single large out-degree vertices to form a tree structure. In this way it may allow out-edge operations to be hierarchically parallelized. Whereas, Aspen and Terrace, and GraphVine target conventional CPUs and GPUs, respectively, our focus is more geared towards near-memory oriented manycore systems.

We use some of the ideas of hierarchical partitioning and augments with other features, courtesy to our message-driven execution model, to create a vertex data structure called the Recursively Parallel Vertex Object (RPVO). 

\subsection{Recursively Parallel Vertex Object (RPVO)}
The RPVO is constructed from vertex objects that are linked together hierarchy. These vertex objects contain a chunk of edges, called \textit{local edge-list}, stored as pointers to other RPVOs representing other vertices of the graph. The root vertex object serves as the user program's accessible address for the vertex. In addition to containing a chunk of edges, the root vertex also holds program data defined by the user. On the other hand, the remaining vertex objects are referred to as \textit{ghost vertices}, and they only contain chunks of edges and pointers to other ghost vertices within the hierarchy. Figure \ref{fig:datastructure-RPVO} shows RPVO representation of the vertex in Figure \ref{fig:datastructure-vertex}. 

The RPVO representation allows scaling the maximum size of a single vertex object beyond the memory limits of a single compute cell (or a tile, in the parlance of similar works). On top of that it also provides recursive parallelism for vertex operations. The child ghost vertices do not wait for parent ghost vertices. When a ghost vertex relays an \textit{action} to its child ghost vertex on a different CC, the child can start execution as soon resources are available at that CC. In this way, logically all vertices (root and ghost) of the RPVO can execute work in parallel. For example, edge search instead of being $\mathcal{O}(\text{edges})$ now becomes $\mathcal{O}(\log_g \text{depth} \times \textit{local edge-list size})$, where $g$ is the number of ghost vertices per vertex. 

Finally, the RPVO provides graceful mutations to the graph structure. Since vertices and edges are pointers they can be created, deleted, or modified on the fly---very powerful for dynamic graph processing. This is in contrast to more rigid matrix oriented representations such as the Compressed Sparse Row (CSR).

Although, hierarchical tree like structures solve the problem of out-degree load, they do not address in-degree load imbalance and congestion in vertex-centric programs. In other words hierarchy can not solve two problems at the same time. This ultimately degrades performance, especially at large core counts and with graphs having highly skewed in-degree distributions. Further studied in Section \ref{subsec:strong-scaling}. To remedy this and also keep the vertex-centric semantics and structure intact we introduce rhizomes laterally in the hierarchical tree structure. To the best of our knowledge, this is the first time such ideas are used together for scalable vertex-centric graph representation.

\subsection{Rhizome}
In botany, a rhizome is a stem that grows horizontally, can grow roots, and can connect to other roots. In human thought, such as philosophy, psychology and social sciences \cite{rhizomeOxfRef} \cite{deleuze1988thousand}, it is used to describe decentralized systems or organizations that form without following any hierarchical structure (arborescent) or predefined order (trace). We borrow these ideas of rhizome from other disciplines and try to incorporate them in our field of engineering.

The RPVO is a very arborescent structure having a root and a hierarchy in which ghost vertices (containing chunks of out-degree edges) exist. We extend the RPVO by adding a new link called the rhizome-link and join many RPVOs representing a single vertex. These distinct RPVOs have their own named addresses to which the in-edges may point. When the graph is constructed, instead of all in-edges pointing to a single RPVO, the in-degree load is shared among the many RPVOs thus forming a rhizome. These in-edges are not stored inside the rhizome but merely point to it. The in-edges exist inside other vertices as out-edges. Figure \ref{fig:datastructure-rhizome} shows a Rhizomatic-RPVO.

\section{Diffusive Programming Model}\label{sec:programming}

Popular, time-tested programming and computing bulk synchronous models of fork-join for many-core systems have served well for tens to hundreds of cores \cite{beyondstatic2023}. At the scales of thousands of cores and ultra fine-granularity there arises a need for a programming and computing model that provides 1) global parallelism with sophisticated event-driven synchronization constructs to avoid the strangler effect \cite{StanglerEffect}, 2) and the ability to express fine-grain operations for irregular memory and low compute intensity applications such as graph processing.

With the above in mind, we consider an approach which can be called the ``\textit{diffusive programming model}'', where an asynchronous action is sent from a memory locality to another memory locality (target). The memory locality can either be on the same CC or on a different CC. This action can mutate the state of the target locality and can further create new actions (work) at the destination thereby creating a ripple effect. When applied to graph processing it has some similarities to the ``Thinking Like a Vertex'' (TLAV) \cite{TLAV}, in that the action may activate a vertex and in turn the vertex can activate its neighboring vertices by diffusing the action along the out-edges. 

\lstset{
  language=C++,
  basicstyle=\ttfamily\footnotesize,
  keywordstyle=\color{blue}\ttfamily,
  stringstyle=\color{red}\ttfamily,
  commentstyle=\color{green!40!black}\ttfamily,
  morecomment=[l][\color{magenta}]{\#},
  showstringspaces=false,
  breaklines=true,
  %backgroundcolor=\color{gray!10},    % Background color
  numbers=left,
  numberstyle=\tiny\color{gray},
  numbersep=2pt,
  upquote=true,
}

\begin{lstlisting}[language=C++,caption=Typical call to a diffusive computation,label=lst:main,xleftmargin=0.3cm]
void main() {
  AMCCA_Device dev = /* Initialize the device. */
  Graph_Type Graph = /* Create the graph and 
                        allocate on the device. */
  int root = 42;
  int level = 0;
  // Get address of the root vertex.
  AMCCA_Pointer root_addr = Graph.get_addr_at(root);
  // Register the `bfs-action` action with the device.
  AMCCA_REGISTER_ACTION(dev, BFS_Action, "bfs-action");
  // Create the action object. 
  // BFS_Action is registered as handler to bfs-action.
  AMCCA_Action bfs_action = AMCCA_Action(BFS_Action,
                                         root_addr, 
                                         level);
  // Create a termintor object that handles 
  // termination detection for the diffusion.
  AMCCA_Terminator terminator = AMCCA_Terminator();                                       
  // Germinate the action on the compute cell 
  // where root_addr exists.
  dev.germinate_action(bfs_action);
  // Diffuse and wait on the terminator.
  dev.run(terminator);
}
\end{lstlisting}

Since the computation progresses asynchronously and has the potential to reactivate a previous node in the execution graph, this means that there is no predetermined computational dependency graph \cite{dependencegraph1981} or one that is determined at runtime \cite{dagbasedruntime2013}. In asynchronous graph processing, there is no dependency graph or the Directed Acyclic Graph (DAG) of the execution. It is primarily because of arbitrary structure of the input data graph with arbitrary weights, whereby the execution (program) relies on some arbitrary invariant to advance \cite{synchavoidJesun2018}. As a result, such computations present an extra challenge of detecting when the execution is complete, known as the Termination Detection Problem (TDP). This problem does not manifest in formulations of the BSP counterparts. Termination detection can be implemented in software by techniques such as Dijkstra–Scholten algorithm \cite{terminationdetection1980} that keep track of actions in the system that are yet to be executed by creating implicit spanning tree of the execution. Software techniques come with overheads of extra acknowledgement messages. Hardware techniques include simple hierarchical signaling mechanisms \cite{harwareTerminationDetection} that relay the aggregate idle status of cores to the host. In this paper we assume hardware signalling.

Listing \ref{lst:main} shows a typical call to an AM-CCA diffusive program in a manner of an accelerator. It solves the Breadth First Search (BFS) using the \langoperator{bfs-action} of Listing \ref{lst:cca-bfs}.

\subsection{Synchronization}

The diffusive programming model allows a decentralized, event-driven, and asynchronous execution regime. As such the use of conventional barrier and blocking oriented synchronization methods will be self-defeating. Methods and techniques of synchronization called Local Control Objects (LCOs), that are argued in ParalleX \cite{paralleX2009}\cite{paralleX2007} and used in HPX \cite{SurveyAMT2017}, preserve global parallelism and fine-granularity. These include use of objects such as \textit{futures} in a way that enable bypassing dependencies in executions as much as possible until the result is needed. In our design actions run to completion (and can not be blocked or preempted), as such the hardware resources of a compute cell will always be open to consume any work that may be available. Control can be transferred back with the use of continuations setting the future. In this paper, we use the \textit{AND Gate LCO}, which is more expressive, having its own trigger-action. An \textit{AND Gate LCO} locally executes its trigger-action when its value is set $N$ number of times. Its use in providing rhizome consistency is discussed in Section \ref{subsec:rhizome-consistency}.

\begin{comment}
\subsection{Continuation}
\todo{Generally write about continuation and specially in the context of CCA.}
One use case is in knowledge graphs where new knowledge is learned and inserted as edges. Consider the syllogism: ``All men are mortal. Socrates is a man. Therefore, Socrates is mortal''. This new edge, \textit{Socrates \textbf{is} mortal}, is dynamically created and inserted in the graph.
\todo{How our use is novel than conventional HPX?}
\end{comment}

\section{Language Constructs and the Runtime}\label{sec:language-runtime}
Using examples of BFS and Page Rank, we introduce our key design ideas related to language constructs and Runtime that aiding in writing actions that mutate the state of rhizomes to perform graph processing. Listing \ref{lst:cca-bfs} shows an action called \langoperator{bfs-action} that traverses the graph to perform BFS. The operand \texttt{v} is the memory address of the vertex on which this action is invoked, and \texttt{lvl} is the BFS level sent from a neighboring vertex. Details of the vertex type are shown in Listing \ref{lst:data-structure-edge} and Listing \ref{lst:data-structure-vertex}.

\lstset{
    language=Racket,
    basicstyle=\ttfamily\scriptsize, % \ttfamily\footnotesize,
    % keywordstyle=\color{blue}\ttfamily,
    stringstyle=\color{red}\ttfamily,
    % commentstyle=\color{orange!80!black}\ttfamily,
    commentstyle=\color{green!40!black}\ttfamily,
    morecomment=[l][\color{magenta}]{\#},
    showstringspaces=false,
    breaklines=true,
    %backgroundcolor=\color{gray!5},
    numbers=left,
    numberstyle=\tiny\color{gray},
    numbersep=2pt,
    upquote=true,
    % frame=single,
}

%\lstinputlisting[language=Racket,caption=Vertex-Centric Data Structure,label=lst:data-structure,xleftmargin=0.3cm]{Codes/data_structure.rkt}

\begin{figure}[H]
  \begin{minipage}[l]{0.19\textwidth}
\begin{lstlisting}[language=Racket,caption=Edge type,label=lst:data-structure-edge,xleftmargin=0.3cm]
; edge type
(struct edge 
  ([address : Pointer] 
   [weight  : Integer]))
\end{lstlisting}
  \end{minipage}\hfill
  \begin{minipage}[l]{0.25\textwidth}
\begin{lstlisting}[language=Racket,caption=Vertex-Centric Data Structure,label=lst:data-structure-vertex,xleftmargin=0.3cm]
; vertex type for BFS
(struct vertex 
  ([id    : Integer] 
   [level : Integer]
   [edges : (Vector edge)]))
\end{lstlisting}
    %\label{fig:cca-chip-memory-controller}
  \end{minipage}
\end{figure}

\begin{lstlisting}[language=Racket,caption=Breadth first search action.,label=lst:cca-bfs,xleftmargin=0.3cm]
;; Breadth First Search Action
(define bfs-action ;; v: target vertex, lvl: level in
  (lambda ([v : (Pointer vertex)] [lvl : Integer])
    (if (> (vertex-level v) lvl)
      (begin
        ;; Perform work.
        (set-vertex-level! v lvl) 
        ;; Diffusion occurs.
        (inform-neighbors (vertex-edges v) (+ lvl 1)) 
      ))))
\end{lstlisting}

\begin{comment}
%\setlength{\intextsep}{1pt} % Adjust vertical spacing between text and floats
\begin{wrapfigure}[8]{l}{0.57\textwidth}
\begin{lstlisting}[language=Racket,caption=BFS Action written in an extended Scheme like language to write asynchronous vertex programs.,label=lst:cca-bfs,xleftmargin=0.1cm]
;; Breadth First Search Action
(define bfs-action ;; v: target vertex, lvl: level in
  (lambda ([v : (Pointer vertex)] [lvl : Integer])
    (if (> (vertex-level v) lvl)
      (begin
        ;; Perform work.
        (set-vertex-level! v lvl) 
        ;; Diffusion occurs.
        (inform-neighbors (vertex-edges v) (+ lvl 1)) 
      ))))
\end{lstlisting}
\end{wrapfigure}
\end{comment}

\begin{lstlisting}[language=Racket,caption=A convenient function to send the new BFS level to neighbors.,label=lst:cca-bfs-inform-neighbors,xleftmargin=0.3cm]
;; Function to send the new BFS level along out-edges
(define inform-neighbors
  (lambda ([edges : (Vector edge)] [lvl : Integer])
    (for-each 
     (lambda (e)
       ;; Get address of vertex along edge e.
       (let ([addr (edge-address e)])
         ;; Send action along edge e.
         (propagate bfs-action (list addr lvl))))
     edges)
  ))
\end{lstlisting}

Listing \ref{lst:cca-bfs-closure} shows the same \langoperator{bfs-action} but using new keywords of \langkeyword{predicate} and \langkeyword{diffuse} that aid in writing actions. These keywords enable the compiler to generate code that the Runtime can use to optimally schedule compute tasks or network operations so as to avoid resources being starved. The \langkeyword{predicate} checks, based on the vertex object's local state and the incoming message payload, whether to activate the vertex and perform work. Think of multiple \langoperator{bfs-action}s received for a single vertex object. Among these actions, only one may contain a better path to the solution based on some heuristic, in this case, the \texttt{lvl} contained in its predicate. This particular action subsumes the work of all other actions, which are then discarded. Using the \langkeyword{predicate} keyword, this check is exposed to the Runtime, which can prune or schedule the action without invoking it. When active, the vertex performs work and then may go into the \langkeyword{diffuse} mode to create new actions that are sent along its out-edges using \langkeyword{propagate} (See Listing \ref{lst:cca-bfs-inform-neighbors}). The programming model assumes an action, with its work and diffusion, to be an atomic unit of execution. Under the hood, the compiler and runtime system work together to separate compute work from diffusion.

Using the \langkeyword{diffuse} keyword, the network operations of \langkeyword{propagate} are detangled from compute operations in the action. This is accomplished by the compiler converting \langkeyword{diffuse} into a closure, which has its \langkeyword{predicate} and body, line \texttt{\#$9$} and \texttt{\#$10$}, respectively in Listing \ref{lst:cca-bfs-closure}. The design also involves adding a second work queue called the \textit{diffuse queue} and this newly created closure is enqueued into it as an anonymous action. It will be executed at some later time by the runtime after the current action finishes. Think about it as lazy (or deferred) evaluation. Although, it can be evaluated immediately, we defer it on purpose. The rationale is two-fold: firstly, it allows actions to be executed without being mechanically tied to their diffusion (although they are logically tied), thereby preventing the computation from blocking on network operations. Secondly, it allows the filtering out of diffusions at a later time during execution when newer actions arrive with better solutions, thereby also subsuming (pruning) the diffusions. Such pruning of the diffusions is helpful for applications, like BFS and SSSP, that monotonically arrive towards a solution.

Furthermore, for such search applications, if there were a single queue (per compute cell) and the current task were blocked on a network operation, then the queue would rapidly gets full as new actions are injected from neighboring compute cells. The computation may deadlock due to the queue being full unless complex deadlock avoidance methods are implemented, which will be very difficult especially considering \textit{fire-and-forget} semantics. 

\begin{comment}
    Section \todo{?} shows performance improvements using this dual queue model against a single queue model, especially showcasing the reduction in the average number of times a vertex (re)activate. Because as diffusions are pruned they don't cause wasted activation of vertices.
\end{comment}

\begin{lstlisting}[language=Racket,caption={\langkeyword{predicate} allows the Runtime to peek into the action and check whether to invoke it or prune it. \langkeyword{diffuse} becomes a closure and is detangled from the action's work, thereby freeing the compute cell from ever stalling an action if the network were to be congested. This helps to keep the compute cell executing actions even when the network is stalled and \langkeyword{propagate}s can not progress.},label=lst:cca-bfs-closure,xleftmargin=0.3cm]
;; Breadth First Search Action
(define bfs-action ;; v: target vertex, lvl: level in
  (lambda ([v : (Pointer vertex)] [lvl : Integer])
    (predicate (> (vertex-level v) lvl)
      (begin
        (set-vertex-level! v lvl) ;; Perform work.
        (diffuse ;; Diffusion occurs with its own
          ;; predicate for pruning.
          (predicate (eq? (vertex-level v) lvl)
            (inform-neighbors (vertex-edges v) (+ lvl 1)))
      ))))

;; The body of `diffuse` is enclosed within a closure
;; that is dispatched to the runtime and enqueued into
;; the diffuse queue of the Compute Cell.
(define diffuse
  (lambda (body)
    (enqueue! sys-diffuse-queue (lambda () (body)))))
              
\end{lstlisting}

In this paper, we restrict the scope of the language to be statically typed and compiled. A dynamically typed and interpreted language will bring opportunities of more expressiveness and runtime optimizations. In particular, the ability to learn symbolic rules and logical relationships from the graph data and based on that new actions expressed and interpreted at runtime. This endeavor is deferred for future work.

\subsection{Rhizome Consistency}\label{subsec:rhizome-consistency}

\begin{figure}
  \centering
  \includegraphics[width=1.0\linewidth]{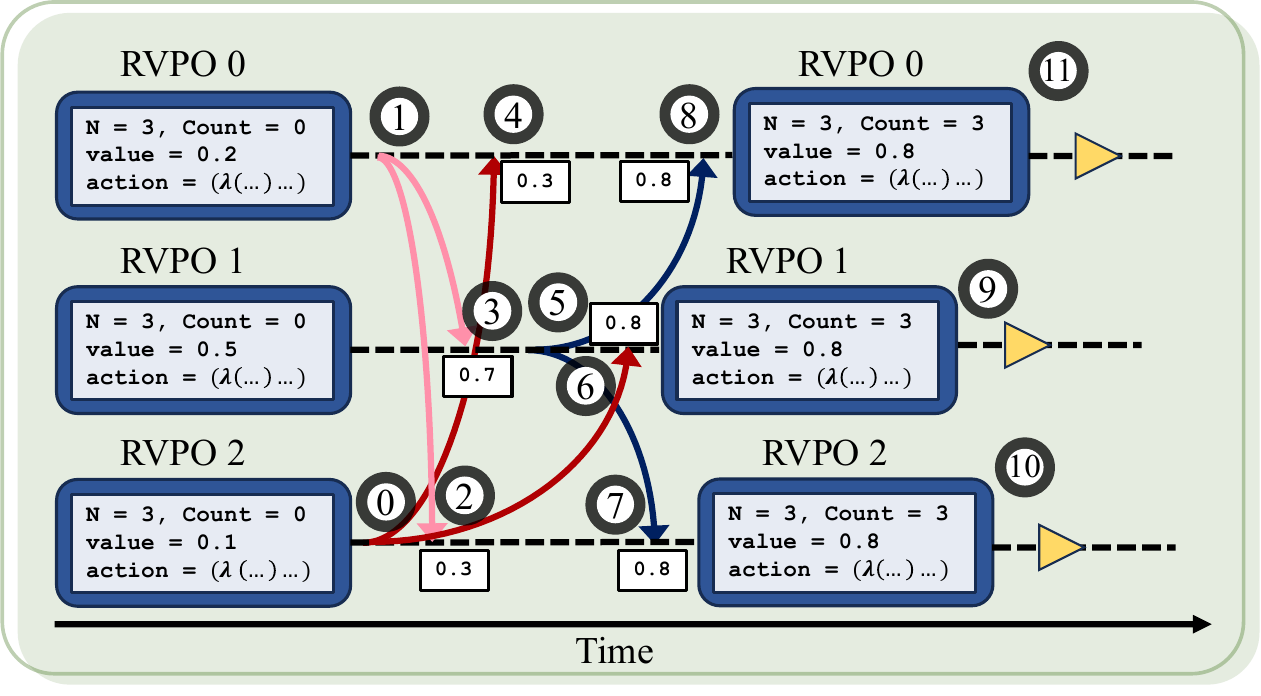}
  %\captionsetup{skip=0pt} % Reduce the space between figure and caption
  \caption{\texttt{score : (\textcolor{violet!80!black}{AND Float})}, an \textit{AND Gate LCO} of \langkeyword{Float} type, as an exemplar shows the internal state of the \textit{AND Gate LCO} object (per RPVO) as it is being used to provide rhizome consistency for Page Rank score for any given single vertex. \protect\poolball{0}\protect\poolball{1}\protect\poolball{5} RPVOs send out their score over the rhizome-link. \protect\poolball{2}\protect\poolball{3}\protect\poolball{4}\protect\poolball{6}\protect\poolball{7}\protect\poolball{8} RPVOs receive score from other RPVOs over the rhizome-link. \protect\poolball{9}\protect\poolballtiny{10}\protect\poolballtiny{11} rhizome has been collapsed, the associated action is triggered locally at each RPVO, and the \texttt{score} AND Gate is reset.}
\label{fig:rhizome-collapse}
\end{figure}

Since the rhizomes are distributed and act independently, their local state may not be globally consistent. In our BFS example, it is the \texttt{level} that may be different for each rhizome. To ensure consistency, in the \langkeyword{diffuse} section, the compiler/user injects \langkeyword{propagate} along the rhizome-links to send \langoperator{bfs-action} with the received operand \texttt{lvl}. For other applications, this can be more involved. For example, Page Rank, where the score is updated each iteration when the total actions received equal the total inbound degree. It will require an all-reduce of the page rank score for that iteration by using \langkeyword{propagate} along rhizome-links. Figure \ref{fig:rhizome-collapse} shows the state of rhizomes for a single vertex and how the \textit{AND Gate LCO}s are used to synchronize its rhizomes.

In Listing \ref{lst:rhizome-collapse}, we introduce a primitive that synchronizes rhizomes to have the correct global view of vertex data. The primitive \langoperator{rhizome-collapse} takes an operator, an LCO, and a trigger action. The operator is applied to the LCO, and when the LCO is set, it locally invokes the given trigger action.

\begin{lstlisting}[language=Racket,caption={Asynchronously synchronize the rhizomes using the \textit{AND Gate LCO}. The compiler converts any type defined/annotated by the property \textcolor{blue}{\texttt{\#:rhizome-shared}} into an \textit{AND Gate LCO} of that type. For BFS, it is the \texttt{level} as shown in Listing \ref{lst:data-structure-rhizome}. For Page Rank, it is the score as shown in \ref{cca-pagerank-closure-rhizome}.},label=lst:rhizome-collapse,xleftmargin=0.3cm]
;; Rhizome consistency
(rhizome-collapse (op LCO) trigger-action)

;; BFS: Broadcast the new level to rhizomes. Immediately
;; reset the `level` AND Gate LCO to execute the trigger 
;; that diffuses the new level to neighbors.
(rhizome-collapse (bcast (vertex-level v))
                  (lambda () (diffuse the new bfs...)))

;; Page Rank: All reduce the score from all rhizomes.
;; When the AND Gate LCO is set, meaning accumulates
;; all values, it triggers the provided trigger action.
(rhizome-collapse (+ (vertex-score v))
                  (lambda ()
                    (vertex-set-score! 
                      (+ (/ (- 1 damping) ...) ...))))
      
\end{lstlisting}

\begin{lstlisting}[language=Racket,caption={Use of the property \textcolor{blue}{\texttt{\#:rhizome-shared}} in definition of a vertex structure.},label=lst:data-structure-rhizome,xleftmargin=0.3cm]
; vertex type for BFS using Rhizomes
(struct vertex 
  ([id    : Integer] 
   [level : Integer] #:rhizome-shared
   [edges : (Vector edge)]))
\end{lstlisting}

For example, the same vertex type of Listing \ref{lst:data-structure-vertex} but now using rhizome is shown in Listing \ref{cca-bfs-closure-rhizome}.
\begin{lstlisting}[language=Racket,caption={BFS action using rhizome},label=cca-bfs-closure-rhizome,xleftmargin=0.3cm]
;; Breadth First Search Action with Rhizome
(define bfs-action ;; v: target vertex, lvl: level in
  (lambda ([v : (Pointer vertex)] [lvl : Integer])
    (predicate (> (vertex-level v) lvl)
      (rhizome-collapse (bcast (vertex-level v))
          (lambda () ;; trigger-action for rhizome-collapse
            (diffuse ;; Diffusion occurs with its own
              ;; predicate for pruning.
              (predicate (eq? (vertex-level v) lvl)
                (inform-neighbors (vertex-edges v) (+ lvl 1)))
            ))))))
\end{lstlisting}

\begin{lstlisting}[language=Racket,caption={Page Rank action using rhizome. Some details are left out for brevity.},label=cca-pagerank-closure-rhizome,xleftmargin=0.3cm]
; vertex type for Page Rank
(struct vertex ([id             : Integer]
                [msg-count      : Integer]
                [score          : Float] #:rhizome-shared
                [edge           : (Vector edge)]))

;; Page Rank Action
(define page-rank-action ;; v: target vertex, score: score in
  (lambda ([v : (Pointer vertex)] [score : Float])
    (predicate (#t) ;; Always true.
      (begin 
        ;; If this iteration has not diffused then diffuse it.  
        (if (eq? (vertex-msg-count v) 0)
          (let ([score-to-send 
                (/ (vertex-score v) (vertex-out-degree v))])
            (begin
              (diffuse ;; Diffusion occurs.
                (predicate (#t) ;; Always true.
                  (inform-score-to-neighbors
                    (vertex-edges v) score-to-send)))
              ;; Reset the score to 0.
              (set-vertex-score! v 0))))
        ;; Perform work. Add the received score.
        (set-vertex-score! v (+ (vertex-score v) score))
        ;; Update message received count.
        (set-vertex-msg-count! v (+ (vertex-msg-count v) 1))         
        ;; Update score if last message of this iteration.                                           
        (if (eq? (vertex-msg-count v) (vertex-in-degree v))
          ;; Make the rhizomes consistent, i.e. allreduce.
          ;; Then trigger the action that updates score.
          (rhizome-collapse (+ (vertex-score v))
            (lambda ()
              (set-vertex-score! 
                (+ (/ (- 1 damping) total-vertices)
                   (* damping (vertex-score v)))))))
      ))))
\end{lstlisting}

\section{Experiments}
The purpose of experiments is to validate the concepts presented in this paper focusing on the aspects of very fine-grain dynamic and adaptive computations, and to understand the performance behavior of these techniques at varying configurations.

\subsection{Methodology}
We developed a simulator using C++ to design and deploy the presented asynchronous message-driven computations and structures, and to understand the runtime behavior of varying AM-CCA configurations, applications, and input datasets. The simulator is high-level enough to be programmed using the diffusive programming model described in Section \ref{sec:programming} and yet low-level enough to simulate individual message movements between CCs. In a single simulation cycle, a message can traverse one hop from one CC to a neighboring CC. We make this assumption since AM-CCA channel links are $256$ bit wide and can easily send the small \textit{messages} of our tested applications in a single flit cycle. Simultaneously, a single CC, can perform either of the two operations: 1) a computing instruction, which is contained in the predicate resolution and work in the user application action, or 2) the creation and staging of a new message when an instance of \langkeyword{propagate} is called. It means that BFS and SSSP actions take $2$--$3$ cycles of compute, whereas Page Rank action takes anywhere from $3$--$70$ cycles of compute. When their diffusions are executed they in turn take cycles proportional to the amount of \textit{local edge-list size}. The simulator is available at \cite{ccasimulator:online}.

\textbf{Energy Cost Model:} The simulation cost model assumes 7nm CMOS with execution logic complexity comparable to embedded RISC-V variants such as zero\_riscy or SiFive using $13.5K$ gates or less. It is supplemented by non-pipelined FPU in $50K$ transistors capable of performing four basic operations. Data memory is comprised of SRAM with leakage power and $64$-bit word access energies as described in \cite{sram7nm2020}. Finally, two NoC variants are evaluated: Cartesian Mesh and 2D Torus-Mesh, with the latter consuming $50\%$ more resources \cite{Dalorex2023}. The total energy to execute an application is a sum of energies required to traverse the network by all emitted messages, SRAM access and leakage, and execution of actions carried by the messages. More details provided in the artifact evaluation.

\textbf{Routing:} The simulator employs turn-restricted routing that is deadlock free and always traverses the minimal path between source and destination \cite{TurnRestricted1992}. Due to the Torus-Mesh network, virtual channels are added to guarantee deadlock freedom \cite{Torus-Routing}, which act as the distance class \cite{DallyNetworksBook2004}. With every new turn the message changes its virtual channel. Owing to its simplicity, the turn-restricted routing can be implemented without the need for complex circuitry and algorithms in the CCs. These choices serve as a straightforward starting point in our exploration of the concepts and designs presented in this paper.

\textbf{Applications:} Using the simulator, we implemented graph BFS, SSSP, and Page Rank. The application are fully asynchronous meaning there is no waiting on the frontier to be explored before moving to the next frontier. Furthermore, there is no meaning of the conventional frontier. The computation is decentralized and vertices asynchronously explore the search space. We verify the results for correctness against known results found using NetworkX \cite{NetworkX2008}. 

\textbf{Datasets:} We perform our experiments using a combination of synthetic graphs, which include RMAT-18 (generated using PaRMAT \cite{PaRMAT} with  $a=0.45$, $b=0.25$, and $c=0.15$) and Erdős-Rényi (generated using NetworkX), as well as real-world graphs. Table \ref{tab:graphdetails} provides details of the graph datasets used in our static graph experiments, containing key insights such as average SSSP length, and degree distributions, which have implications on application, data structure, and system behavior. To make the SSSP meaningful, random weights are assigned to the edges of both synthetic and real-world graphs.

\begin{table*}
  \caption{Details of the Input Data Graphs}
  \label{tab:graphdetails}
  \centering  
  \begin{tabular}{|c|c|c|c|c|c|c|c|c|c|c|c|c|}
    \toprule
    \hline
    & & & 
    \multicolumn{2}{c}{\multirow{1}{*}{\textbf{SSSP}}} &  \multicolumn{4}{|c|}{\multirow{1}{*}{}} & \multicolumn{4}{|c|}{\multirow{1}{*}{}}\\
    & & \textbf{Directed} & \multicolumn{2}{c}{\textbf{Length ($l$)}} & \multicolumn{4}{|c|}{\textbf{In Degrees ($k_{in}$)}} & \multicolumn{4}{|c|}{\textbf{Out Degrees ($k_{out}$)}} \\
    \cline{4-13}
    \textbf{Graph Name} & \textbf{Vertices} & \textbf{Edges} & $\mu$ & $\sigma$ & $\mu$ & $\sigma$  & max & $<\%,\%tile>$ & $\mu$ & $\sigma$ & max & $<\%,\%tile>$\\
    \hline\hline
    %\midrule
    language: LN & \numtothousand{399130} & \numtomillion{1216334} & \num{7.5} & \num{1.7} & \num{3.0} & \num{3.9} & \num[group-separator={,}]{107} & $<99\%,17>$ & \num{3.0} & \num{20.7} & \numtothousand{11555} & $<99\%,20>$\\
    \hline
    amazon0302: AM & \numtothousand{262111} & \numtomillion{1234877} & \num{8.8} & \num{1.8} & \num{4.7} & \num{5.7} & \num{420} & $<99\%,23>$ & \num{4.7} & \num{0.9} & \num{5} & $<99\%,5>$\\
    \hline
    Erdős-Rényi: E18 & \numtothousand{262144} & \numtomillion{2359296} & \num{4.6} & \num{0.5} & \num{9.0} & \num{3.0} & \num{25} & $<99\%,17>$ & \num{9} & \num{3.0} & \num{25} & $<99\%,17>$\\
    \hline
    RMAT: R18 & \numtothousand{262144} & \numtomillion{4718592} & \num{3.3} & \num{0.5} & \num{18.1} & \num{63.3} & \numtothousand{7536} & $<96\%,87>$ & \num{18.1} & \num{17.6} & \num{488} & $<96\%,55>$\\
    \hline
    %web-Google: WG & \numtothousand{916428} & \numtomillion{5105039} & \num{6.4} & \num{1.7} & \num{5.8} & \num{39.2} & \numtothousand{6326} & $<96\%,23>$ & \num{5.8} & \num{6.5} & \num{456} & $<96\%,19>$\\
    %\hline
    LiveJournal: LJ & \numtomillion{4847571} & \numtomillion{68993773} & - & - & \num{14.2} & \num{43.4} & \numtothousand{13906} & $<98\%,99>$ & \num{14.2} & \num{36.0} & \numtothousand{20293} & $<98\%,98>$\\
    \hline
    Wikipedia: WK & \numtomillion{4206289} & \numtomillion{101311613} & - & - & \num{24.0} & \num{412.9} & \numtothousand{431795} & $<98\%,152>$ & \num{24.0} & \num{47.8} & \numtothousand{8104} & $<98\%,129>$\\
    \hline
    RMAT: R22 & \numtomillion{4194303} & \numtomillion{128311436} & - & - & \num{30.5} & \num{345.6} & \numtothousand{162839} & $<98\%,308>$ & \num{30.5} & \num{345.6} & \numtothousand{162839} & $<98\%,308>$\\
    \hline

    %RMAT: R25 & \numtomillion{33554432} & \numtomillion{659101568} & - & - & \num{19.6} & \num{344.0} & \num[group-separator={,}]{444823} & $<99\%,233>$ & \num{19.6} & \num{344.0} & \num[group-separator={,}]{444823} & $<99\%,233>$\\
    %\hline

    \bottomrule
    \multicolumn{13}{l}{\footnotesize $\mu$ is mean, $\sigma$ is standard deviation, and the pair $<\%,\%tile>$ represents percentile $\%$ and it's value $\%tile$.}\\
    \multicolumn{13}{l}{\footnotesize $l$ is found by averaging SSSP length of a sample of \num{100} vertices. K is thousand, and M is million.}\\
    \multicolumn{13}{l}{\footnotesize R22 is undirected but represented as directed, hence exhibiting symmetry in out-degrees and in-degrees.}\\
  \end{tabular}
\end{table*}

\textbf{Graph Construction:}
\begin{comment}
\subsubsection{Ingestion:}
\todo{Needs re-writing:}
The graph is constructed by first allocating the root RPVO objects on the AM-CCA chip. Once the vertices are allocated and their addresses are known the edges are ingested into the chip by sending a message containing the edge using the \texttt{insert-edge-action} of Listing \ref{lst:cca-insert-edge}. In the case of dynamic graph applications such as Dynamic-BFS, when an edge is inserted, a \texttt{bfs-action} of Listing \ref{lst:cca-bfs-closure} is germinated inside that compute cell. In subsequent cycles, this action will be invoked, performing BFS edge traversal starting from the newly added edge, thereby reusing the prior computation. Fig. \ref{fig:cca-chip-memory-controller} provides further details, especially in the context of on and off-chip data movement.
\end{comment}
%\subsubsection{Rhizome Construction:}
The graph is constructed by first allocating the root RPVO objects on the AM-CCA chip. Once the vertices are allocated and their addresses are known the edges are inserted. When the \textit{local edge-list size} is reached a new ghost vertex is allocated and the edge is inserted into that ghost vertex. For rhizome creation, the highly skewed in-degree vertices of the graph are assigned rhizomes based on:
\begin{enumerate}
  \item \texttt{rpvo\_max}: the desired maximum number of RPVOs per Rhizome.
  \item \texttt{indegree\_max}: max in-degree in the graph.
\end{enumerate}
Equation \ref{eq:cutoff_chunk} uses these variables to derive a cutoff chunk. Whenever an RPVO is pointed to by \texttt{cutoff\_chunk} amount of edges, a new RPVO is created for that vertex and the next \texttt{cutoff\_chunk} edges are pointed to this new RPVO. This process is repeated until \texttt{rpvo\_max} RPVOs are created after which we cycle back to the first RPVO. The reason for the cutoff to be derived from the maximum in-degree of the input graph data is to be consistent in our experimental method for various input graphs. In essence it is not depended on prepossessing the input graph for its degree distribution. It can be a learned constant.

\begin{equation}\label{eq:cutoff_chunk}
\texttt{cutoff\_chunk} = \frac{\texttt{indegree\_max}}{\texttt{rpvo\_max}}
\end{equation}

\begin{figure}
  \centering
  \begin{subfigure}{0.31\linewidth}
    \centering
    \includegraphics[width=\linewidth]{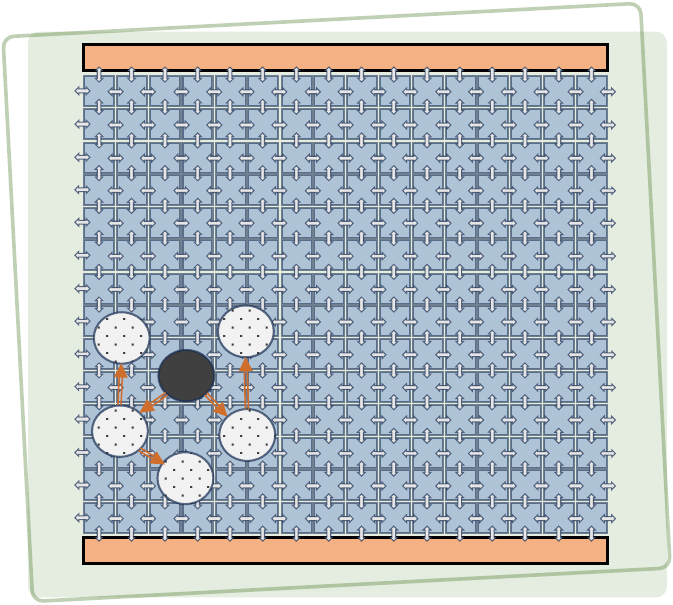}
    \caption{Vicinity Alloc.}
    \label{fig:RPVO-allocator}
  \end{subfigure}
   \begin{subfigure}{0.31\linewidth}
   \centering
   \includegraphics[width=\linewidth]{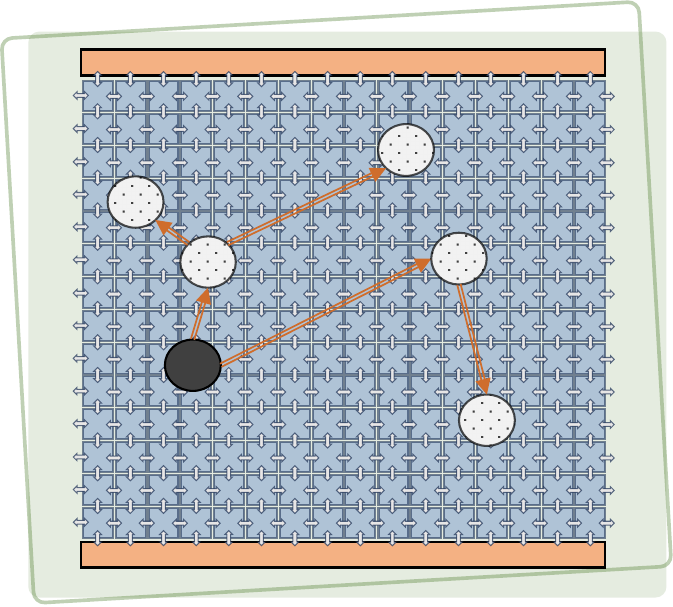}
   \caption{Random Alloc.}
   \label{fig:random-allocator}
  \end{subfigure}
  \begin{subfigure}{0.31\linewidth}
    \centering
    \includegraphics[width=\linewidth]{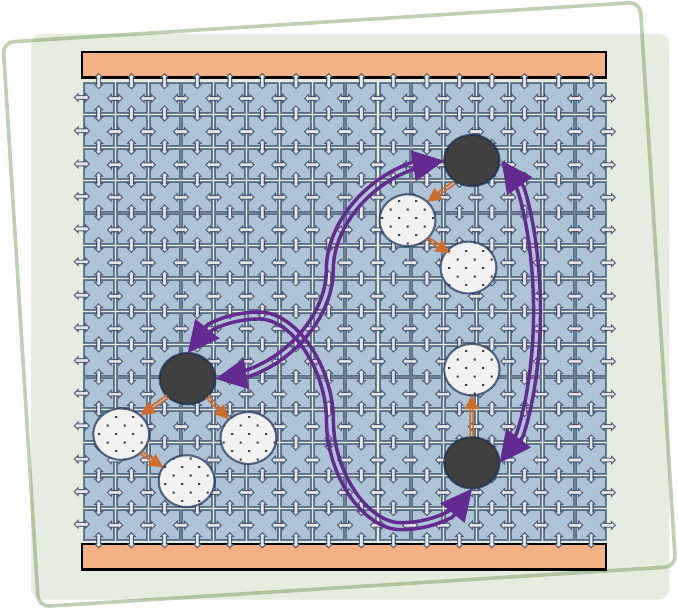}
    \caption{Mixed Alloc.}
    \label{fig:Rhizomes-allocator}
  \end{subfigure}
  \caption{Vertex object allocation policy: (a) Localize ghost vertices in Compute Cells nearby, (b) No regard to locality of ghost vertices, (c) Disperse rhizomes to far away Compute Cells using random allocator while keeping ghost vertices localized using vicinity allocator.}
    \label{fig:allocators-mini}
\end{figure}

\textbf{Affinity of Object Allocation:} We developed two vertex object allocation policies: \textit{Random Allocator} and \textit{Vicinity Allocator}, conceptually shown in Figure \ref{fig:allocators-mini}. As the name implies, the \textit{Random Allocator} randomly allocates vertex objects on any CC across the entire chip, whereas the \textit{Vicinity Allocator} randomly allocates nearby, aiming to reduce the latency of intra-vertex operations when used to allocate ghost vertices.

\begin{comment}
If two or more ghost vertices for a single vertex are allocated on the same CC, it may result in the serialization of recursive (hierarchical) parallelism. To minimize the probability of serialization when using the \textit{Vicinity Allocator}, the radius of allocations, originating from the ghost vertex location, must be large enough to ensure that ghost vertices are allocated in distinct CCs as much as possible.
\end{comment}

To allocate the rhizomes, we use a random allocator to randomly assign the newly created root RPVO of the rhizome to any compute cell on the AM-CCA chip. The hope is that randomness may have allocations distributed across various regions of the chip, thereby avoiding the creation of hot spots in any particular region. Figure \ref{fig:RPVO-allocator} shows ghost vertices of a single RPVO created using vicinity allocator to localize out-degree load and bound the intra-vertex operation latency. Figure \ref{fig:Rhizomes-allocator} shows how a rhizome will be formed by having its RPVOs randomly allocated somewhat far way from each other.

Having the rhizome allocated randomly exhibits some properties of Leslie Valiant's random routing algorithm \cite{valiantRandom}, but the main difference here is that not all actions may eventually be routed to other rhizomes. It is because the rhizomes themselves are the actual destination and not passive proxies. Depending on the application semantics and runtime context, the \langkeyword{predicate} may be false, as is the case with BFS and SSSP, or not all vertex activations require a subsequent action delivery to out-edges or rhizome-links, as described earlier for Page Rank.

\subsection{Diffusion Throttling}
\begin{figure}
  \centering
  \begin{subfigure}{.49\linewidth}
    \centering
    \includegraphics[width=\linewidth]{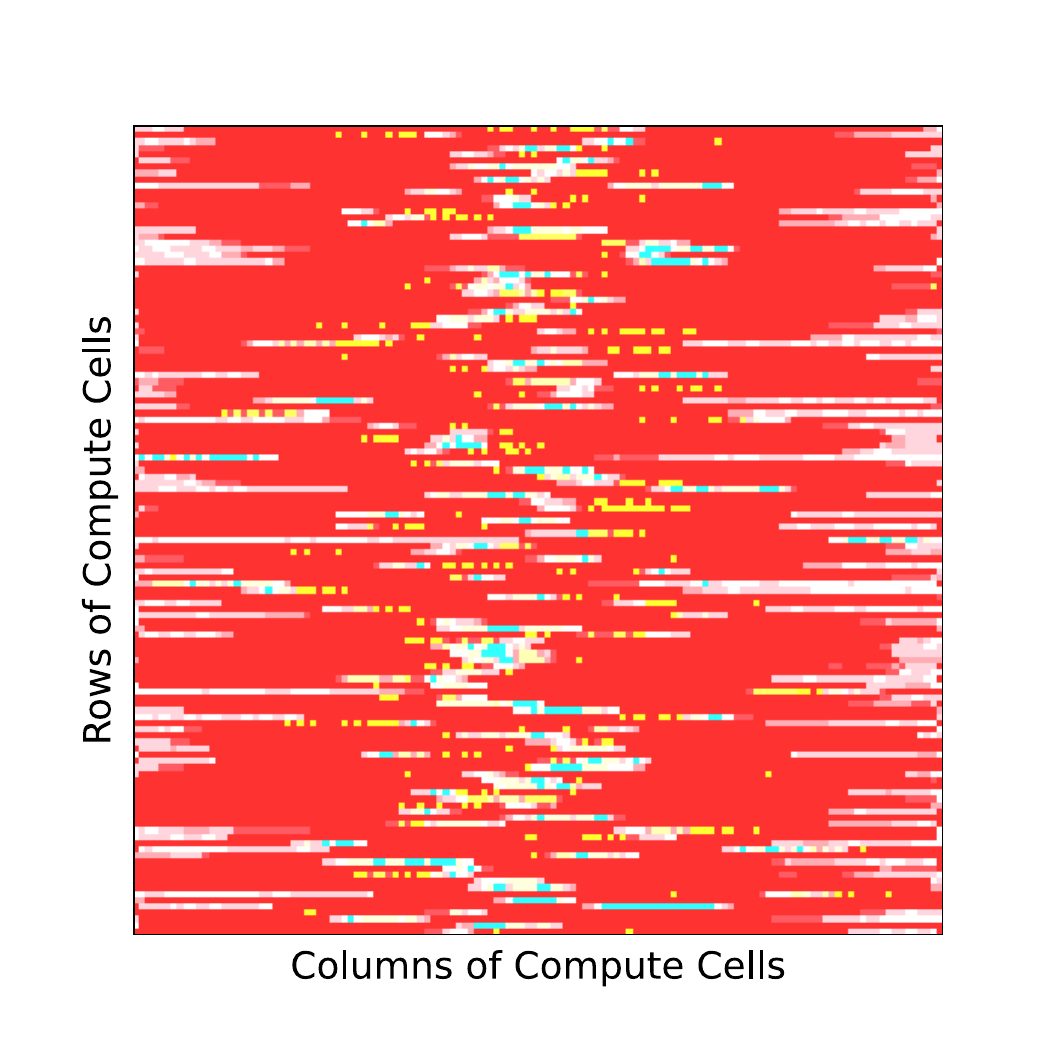}
    \captionsetup{skip=0pt} % Reduce the space between figure and caption
    \caption{Throttle: Off}
    \label{fig:congestion-throttle-off-shuf-on}
  \end{subfigure}
  \hfill
  \centering
  \begin{subfigure}{.49\linewidth}
    \centering
    \includegraphics[width=\linewidth]{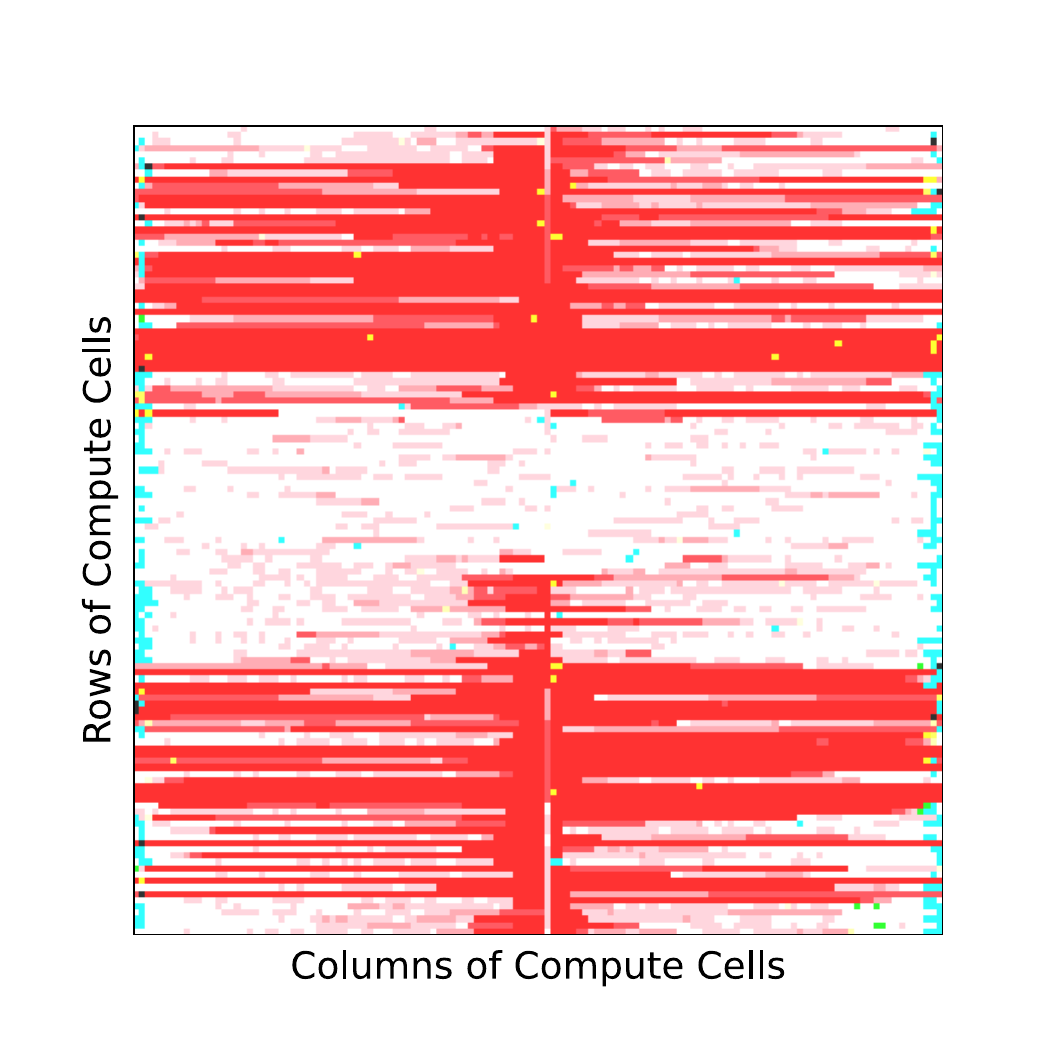}
    \captionsetup{skip=0pt} % Reduce the space between figure and caption
    \caption{Throttle: On}
    \label{fig:congestion-throttle-on-shuf-on}
  \end{subfigure}
  \hfill
  \begin{subfigure}{0.95\linewidth}
    \centering
    \includegraphics[width=\linewidth]{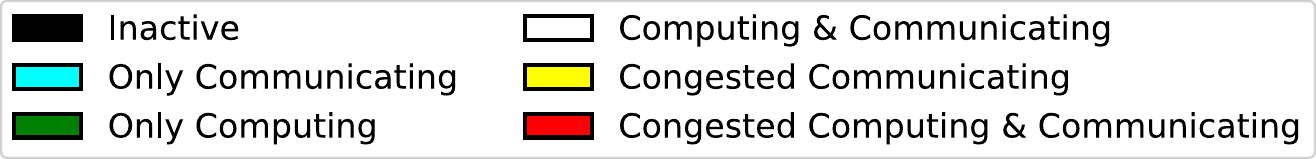}
    \label{fig:congestion-throttle-legend}
  \end{subfigure}
  \captionsetup{skip=0pt} % Reduce the space between figure and caption
  \caption{A moment during the application run showing status per compute cell. There are $128 \times 128$ compute cells with per virtual channel buffer size of $4$ solving the BFS of the RMAT-$18$ graph.}
  \label{fig:congestion-throttle}
\end{figure}

Figure \ref{fig:congestion-throttle} visualizes a moment during execution of BFS traversing the RMAT-$18$ graph. It shows the status of individual CCs, particularly highlighting whether they are congested. We observe that very high ingress of diffusion messages, dictated by the input graph out-degree distribution, if unchecked, quickly creates congestion on the network with \textit{messages} waiting on network channels. This ultimately leads to compute cells being unable to generate new \textit{messages}, thereby stalling the computation. Although, as discussed in Section \ref{sec:language-runtime} regarding pruning, when a computation in diffusion state is stalled, the Runtime overlaps it with either an action execution from the \textit{action queue} or filter passes on \textit{action queue} and \textit{diffuse queue}, that peek into the \langkeyword{predicate} in the hope of work pruning; nevertheless, the network operations of \langkeyword{propagate} remain stalled.

\textbf{Throttling:} We employ a simple mechanism. When a compute cell generates new messages, it first checks for congestion with its immediate neighbors for the previous cycle. Based on congestion, it halts the creation of any new messages for a set period of cycles $T$, in a hope to cool down the network. Equation~\ref{eq:throttling-period} shows the hypotenuse of the chip, which is used as the throttling period $T$. 

\begin{equation}\label{eq:throttling-period}
T = \begin{cases}
      \sqrt{dim_x^2 + dim_y^2} & \text{if Mesh} \\
      \frac{\sqrt{dim_x^2 + dim_y^2}}{2} & \text{if Torus-Mesh}
    \end{cases}
\end{equation}
\begin{comment}
\begin{figure}
  \centering
  \begin{subfigure}{1\linewidth}
    \centering
    \includegraphics[width=\linewidth]{Performance/throttle/contention-chart-th-off.pdf}
    \caption{Cycles spent in congestion: Throttle Off, Shuffle On.}
    \label{fig:contention-chart-throttle-off}
  \end{subfigure}
  \hfill
  \begin{subfigure}{1\linewidth}
    \centering
    \includegraphics[width=\linewidth]{Performance/throttle/contention-chart-th-on.pdf}
    \caption{Cycles spent in congestion: Throttle On, Shuffle On.}
    \label{fig:contention-chart-throttle-on}
  \end{subfigure}
  \caption{Histogram of contention experienced per channel for all compute cells. There are $128 \times 128$ compute cells with per virtual channel buffer size of $4$ solving the Static-BFS of the RMAT-$18$ graph. Predictably, the west and east channels are highly contented due to the X-Y dimension order routing.}
  \label{fig:contention-chart-throttle}
\end{figure}
\end{comment}
When enabled, throttling relieves the message pressure on the network. This can be visually seen in Figure \ref{fig:congestion-throttle-on-shuf-on}. However, some congestion can still be seen, especially in horizontal chunks. This congestion is attributed to the nature of the routing algorithm used, which relies on static horizontal first dimension order routing.

\begin{comment}
Figure \ref{fig:contention-chart-throttle} shows histogram of contention experienced per channel for all compute cells. Due to the horizontal (X-Y) dimension order routing, the west and east channels experience high contention, as depicted in Figure \ref{fig:contention-chart-throttle-off}. Although throttling reduces contention on these channels, as shown in Figure \ref{fig:contention-chart-throttle-on}, there is still room for adaptive routing and less aggressive throttling algorithms that can better utilize available buffers for performance improvements.
\end{comment}

\textbf{Blocking Techniques:} There are blocking methods available to mitigate the issue of congestion, such as message coalescing and message reduction, either in Runtime as in the Active Pebbles system \cite{activepebble2011} or in hardware incarnations as in pipeline architectures such as GraphPulse \cite{GraphPulse2020}. Message coalescing involves grouping the messages before sending at the expense of making the application a little coarser, and message reduction involves staging and processing messages at the sender in a hope to reduce redundant messages aimed at a single receiver. However, we did not choose to implement these blocking methods as the purpose of our experiments is to expose, exploit, and understand unbounded parallelism. Using blocking techniques would limit the amount of instantaneous parallelism and defeat the purpose of this endeavor.

\textbf{Lazy Diffuse as Implicit Reduction:} However, in our design, when the Runtime lazy evaluates a \langkeyword{diffuse} by sending it into the \textit{diffuse queue} to be executed at a later time, it creates an opportunity to prune (reduce) the diffusions. As the \langkeyword{diffuse} has its own \langkeyword{predicate}, which is evaluated at a later time when that \langkeyword{diffuse} is eventually executed, it can be pruned thereby reducing the amount of diffusions. Furthermore, when a \langkeyword{propagate} inside a \langkeyword{diffuse} is blocked on the network due to congestion until its throttling period elapses, the Runtime context switches that diffusion and overlaps it with action execution from the \textit{action queue} or prunes diffusions contained in the \textit{diffuse queue} by peeking into their \langkeyword{predicate}. Therefore, instead of having the compute cell waiting on contention and starve for work, the Runtime ensures that it is occupied by any work that is present in either of the queues without worrying about if any task will block. The actions are now guaranteed to not block and can be overlapped while a \langkeyword{diffuse} is context switched and blocked on the network. This is due to the fact that any code that may block is captured in the \langkeyword{diffuse} clause that is lazy evaluated, and therefore the actions do not have any logical and/or outside physical hardware dependencies. This is because both the following principles hold:
\begin{enumerate}
    \item \textbf{\textit{Principle of logical precedence}}: dependent operations must have completed.
    \item \textbf{\textit{Principle of coincidization}}: dependent data must be available at the location of execution.
\end{enumerate}

\begin{figure}
  \includegraphics[width=\linewidth]{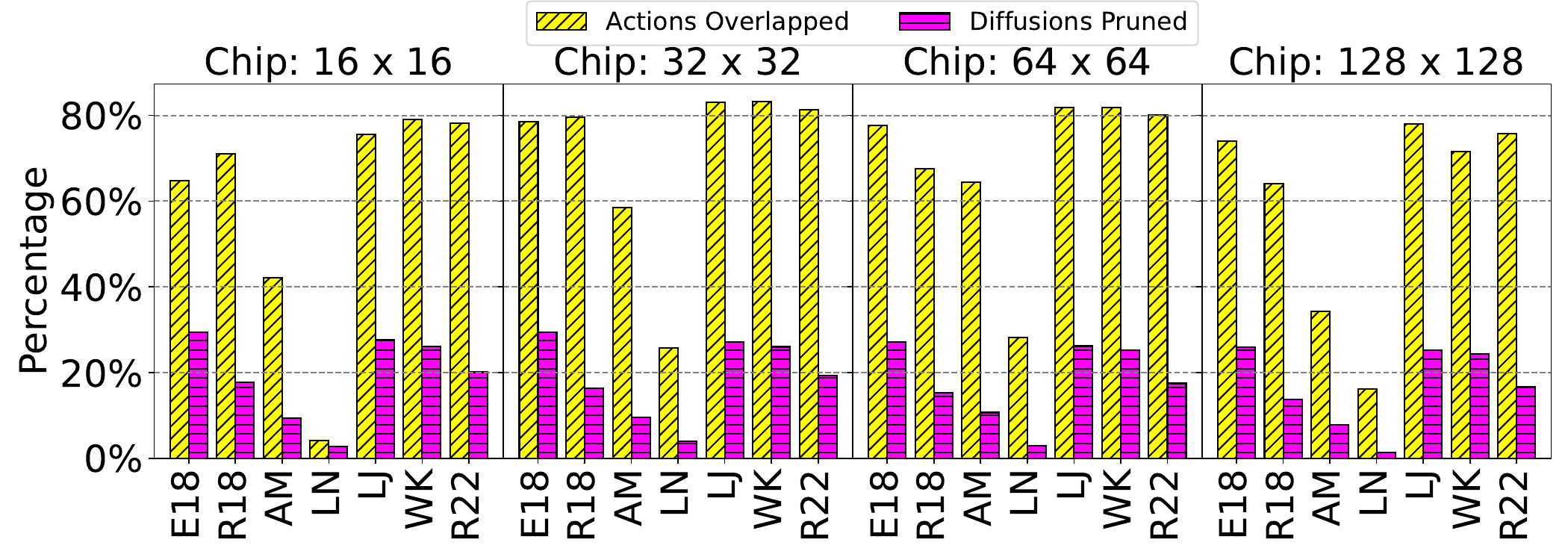}
  \captionsetup{skip=0pt} % Reduce the space between figure and caption
  \caption{Opportunities of lazy diffuse evaluation. Shows the percentage of actions that were overlapped and diffusions in the \textit{diffuse queue} were pruned when a \langkeyword{propagate} was blocked on the network because of congestion. The diffusion were also pruned when they were executed since now they also have their own predicate.}
  \label{fig:lazy-diffuse-performance}
\end{figure}

Figure \ref{fig:lazy-diffuse-performance} shows the percentage of actions overlapped and diffusions pruned for BFS for all datasets and various chip sizes. For monotonically relaxing computations such as our fully asynchronous BFS, not all actions that are created also create diffusions since most of them are false on their \langkeyword{predicate}. In our experiments, across datasets and chip sizes, about 3\%--10\% of the actions perform work (expect AM that had 23\%, E18 had 15\%, and LN had 35\%), whereas the rest of the actions are false on their \langkeyword{predicate}. Of the ones that perform work and create diffusions, not all diffusions may execute since the Runtime prunes them out at a later time when new BFS levels are received and are updated by action executions.

\subsection{Strong Scaling}\label{subsec:strong-scaling}
\begin{figure*}
  \includegraphics[width=\linewidth]{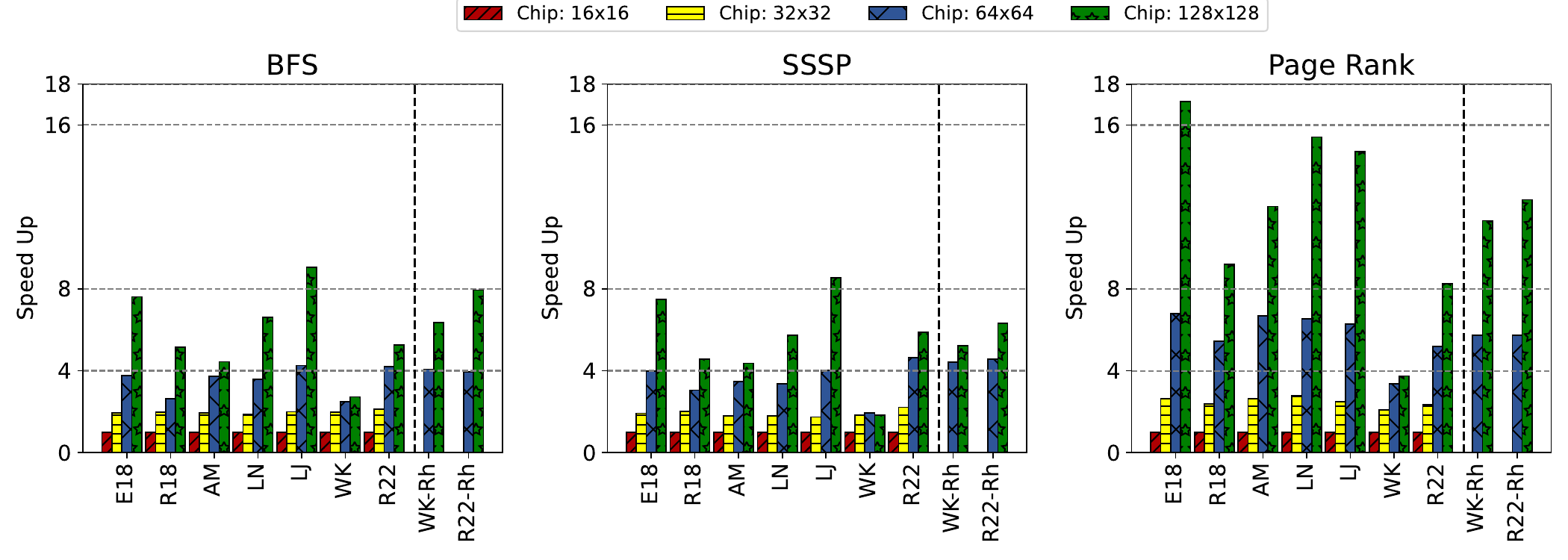}
  \captionsetup{skip=0pt} % Reduce the space between figure and caption
  \caption{Strong scaling using Torus-Mesh network. WK-Rh and R22-Rh use Rhizomatic-RPVO while rest of the data points represent only a simple RPVO (with \texttt{rpvo\_max=1}).}
  \label{fig:strong-scaling}
\end{figure*}

Figure \ref{fig:strong-scaling} shows the strong scaling of BFS, SSSP, and Page Rank running on Torus-Mesh from $256$ compute cells ($16\times16$) to $16,384$ compute cells ($128\times128$). Although the system provides relatively better speed ups for simple RPVO (no rhizomes) but degrades at large compute cell counts with graphs that have highly skewed in-degree distributions. These configurations are the input graphs WK and R22 at chip sizes of $64\times64$ and $128\times128$.

\begin{figure}
  \centering
  \begin{subfigure}{0.24\linewidth}
    \centering
    \includegraphics[width=\linewidth]{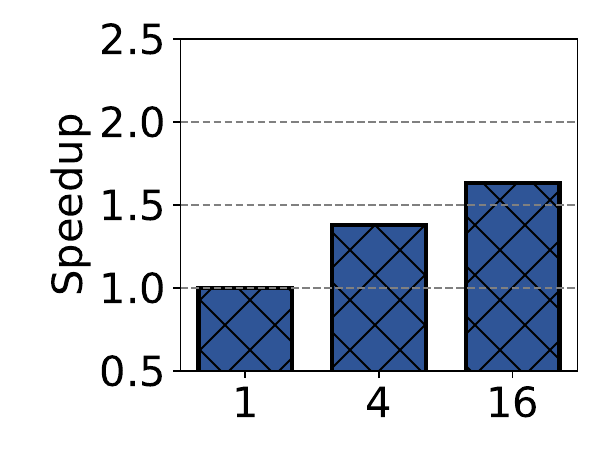}
    \caption{WK on Chip Size: $64\times64$.}
    \label{fig:perf-rhizomes-64-WK}
  \end{subfigure}
  %\hfill
  \begin{subfigure}{0.24\linewidth}
    \centering
    \includegraphics[width=\linewidth]{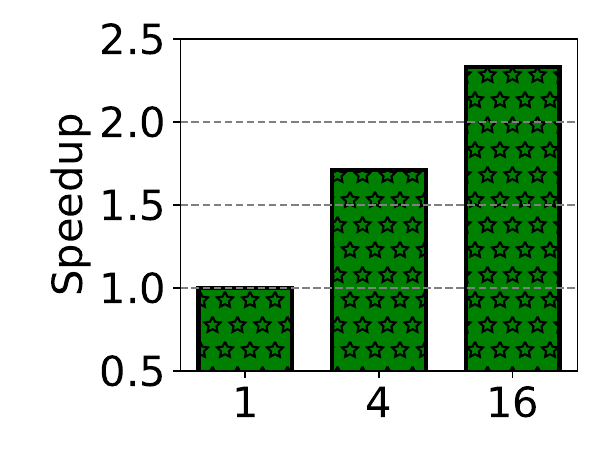}
    \caption{WK on Chip Size: $128\times128$.}
    \label{fig:perf-rhizomes-128-WK}
  \end{subfigure}
  %\hfill
  \begin{subfigure}{0.24\linewidth}
    \centering
    \includegraphics[width=\linewidth]{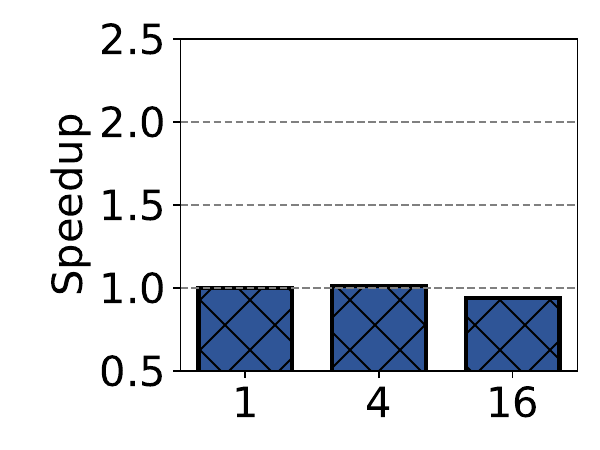}
    \caption{R22 on Chip Size: $64\times64$.}
    \label{fig:perf-rhizomes-64-R22}
  \end{subfigure}
  %\hfill
  \begin{subfigure}{0.24\linewidth}
    \centering
    \includegraphics[width=\linewidth]{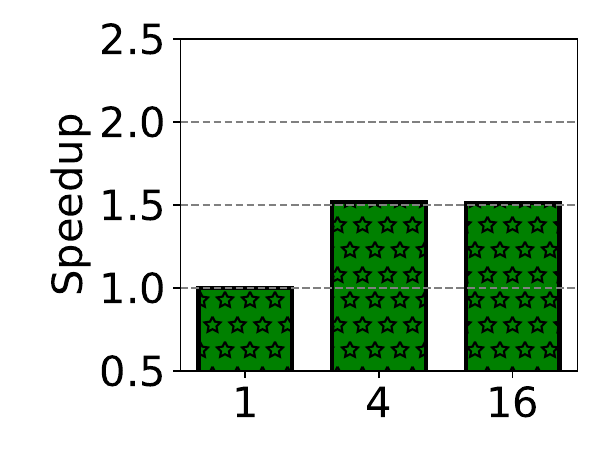}
    \caption{R22 on Chip Size: $128\times128$.}
    \label{fig:perf-rhizomes-128-R22}
  \end{subfigure}
  \caption{Performance of BFS with varying the number of max RPVOs per rhizome. The x-axis represents \texttt{rpvo\_max}.}
  \label{fig:perf-rhizomes}
\end{figure}

\textbf{Deployment of Rhizomes:} We varied the maximum number of rhizomes from $1$, meaning only a single RPVO, to $16$ rhizomes for any single large in-degree vertex. We ran our BFS implementation on the target chip sizes, $64\times64$, and $128\times128$. Figure \ref{fig:perf-rhizomes} shows speed ups against having no rhizomes for graph WK and R22. Only the chip size of $64\times64$ for graph R22 did not scale when using rhizomes. This hints at a relationship of chip sizes for certain graphs sizes and degree distributions at which rhizomes may not provide any advantages. In particular, when we look at the strong scaling in Figure \ref{fig:strong-scaling}, we find that for chip size $64\times64$ BFS operating on R22 graph without rhizomes does have some scalability as compared to the same but with the WK graph. In part, this means that there was not much performance for rhizomes to squeeze any further performance.

\begin{figure}
  \centering
  \begin{subfigure}{.9\linewidth}
    \centering
    \includegraphics[width=\linewidth]{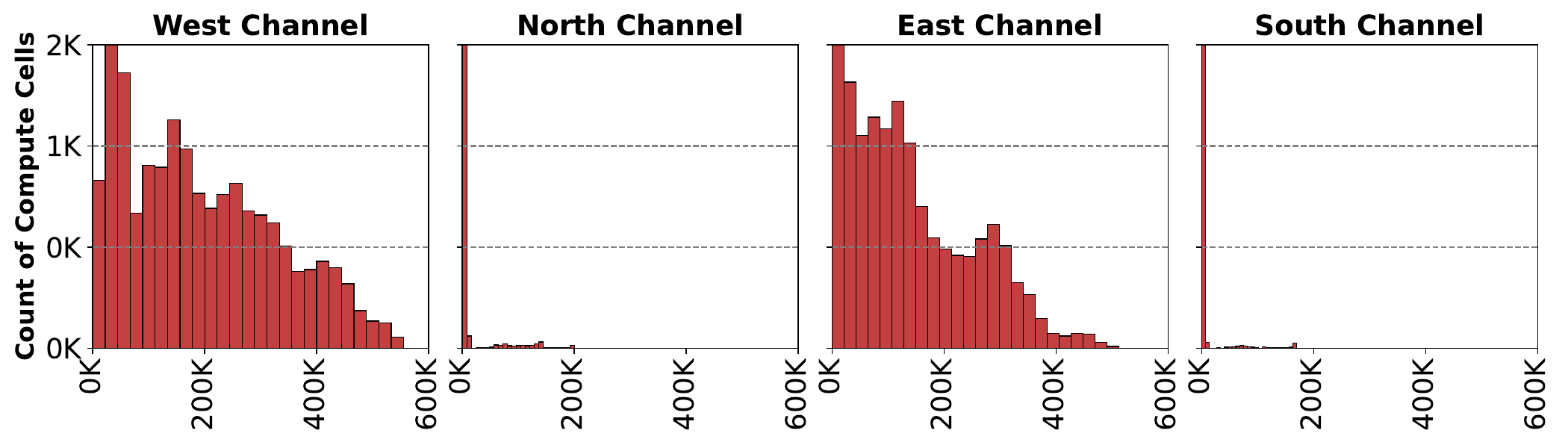}
    \caption{Cycles spent in congestion with $\texttt{rpvo\_max}=1$.}
    \label{fig:contention-chart-rhizome-1}
  \end{subfigure}
  \hfill
  \begin{subfigure}{.9\linewidth}
    \centering
    \includegraphics[width=\linewidth]{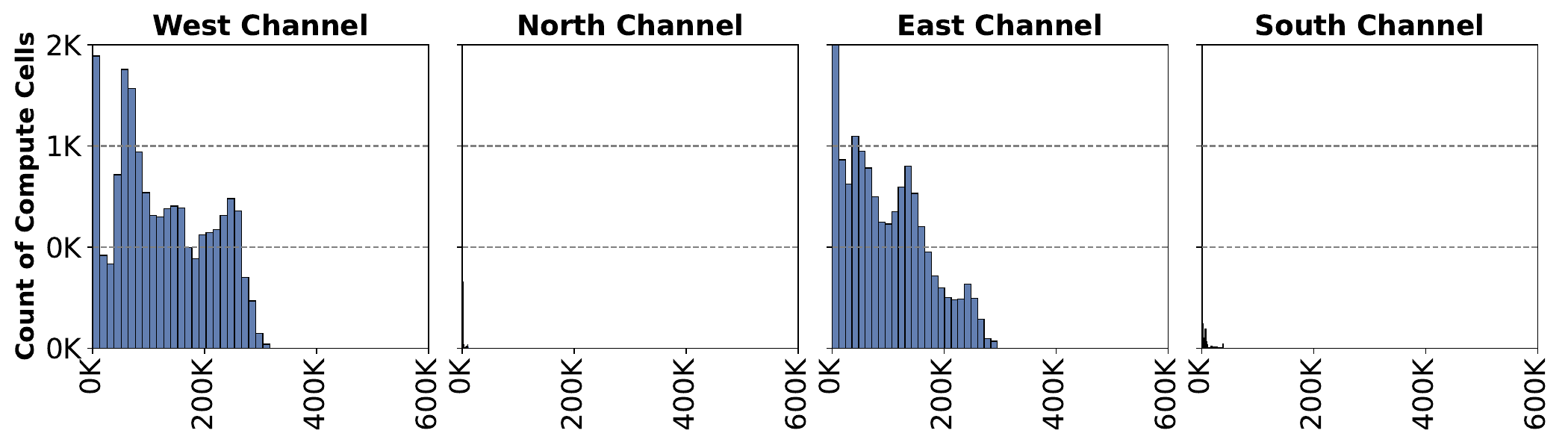}
    \caption{Cycles spent in congestion with $\texttt{rpvo\_max}=16$.}
    \label{fig:contention-chart-rhizome-16}
  \end{subfigure}
  \caption{The introduction of rhizomes lowers contention. Histogram ($bins=25$) of contention experienced per channel for all compute cells for AM-CCA chip size of $128 \times 128$ solving the BFS of the RMAT-$22$ graph. The North and South channels are not as congested due to the static X-Y dimension order routing, which prefers horizontal channels.}
  \label{fig:contention-chart}
\end{figure}

Figure \ref{fig:contention-chart} shows histogram of contention experienced by all compute cells for each channel. It reveals that rhizome performance not only comes from being able to distribution work load of single large vertices but also from distributing the network traffic much evenly thus lowering the congestion. 

\begin{comment}
\subsubsection{Strong Scaling of Rhizomes}\label{subsec:strong-scaling}
Finally, we did strong scaling evaluation of AM-CCA using rhizomatic RPVO and compare it against a simple RPVO containing no rhizomes. Compared to a chip size of $32\times32$ with no rhizomes, Figure \ref{fig:strong-scaling} shows a geomean speed up of $2.08\times$ and $3.75\times$ for chip sizes $64\times64$ and $128\times128$, respectively.
\end{comment}
\subsection{Mesh vs Torus-Mesh}\label{subsec:mesh-vs-torus}
\begin{figure}
  \includegraphics[width=\linewidth]{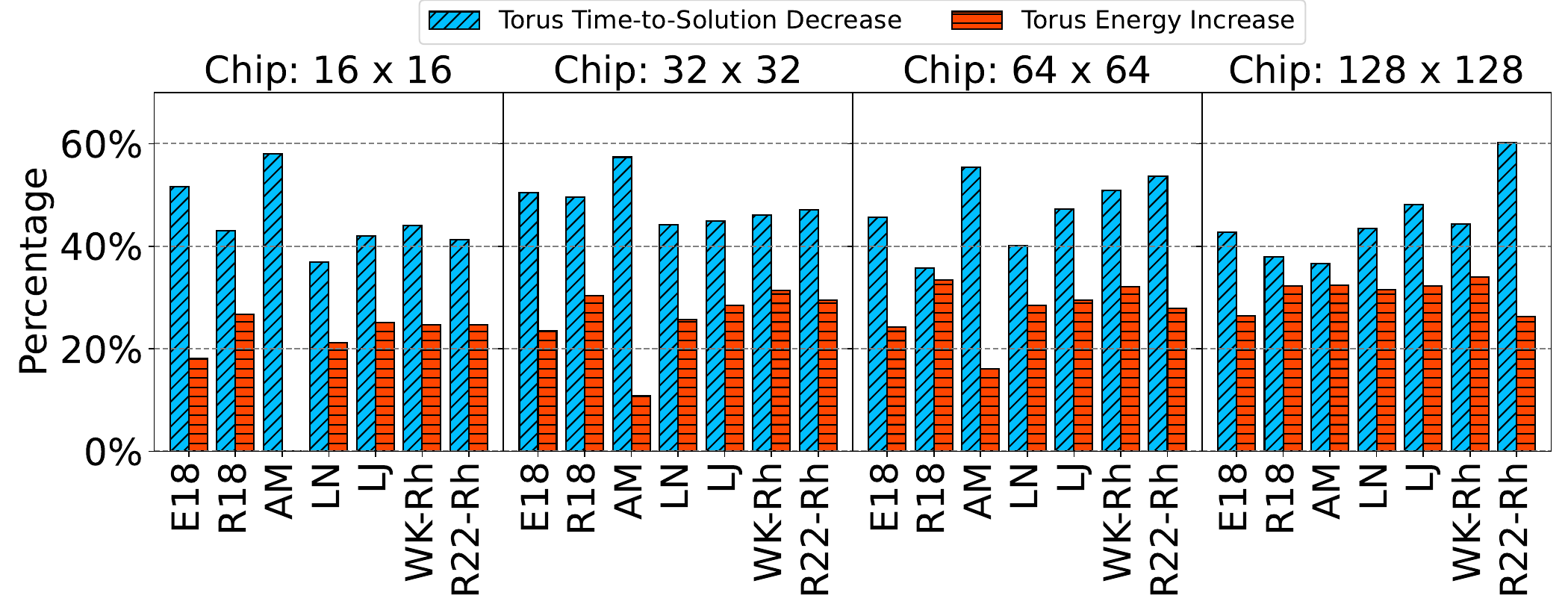}
  \captionsetup{skip=0pt} % Reduce the space between figure and caption
  \caption{BFS performance on Torus-Mesh compared to a simple Mesh.}
  \label{fig:mesh-torus-performance}
\end{figure}

We compared the performance of a pure Mesh vs a Torus-Mesh. Figure \ref{fig:mesh-torus-performance} show the percentage of reduction in time-to-solution and increase in energy consumption for the Torus-Mesh as compared to a pure Mesh. Across datasets and chips sizes there is a geomean $45.9\%$ decrease in time-to-solution and a geomean $26.2\%$ increase in energy cost. 
An anomaly is the energy consumption of $16 \times 16$ Torus-Mesh with the AM graph. The reason is that the AM graph had very few messages propagated due to the small out-degree and coupled with the nature of the $16 \times 16$ Torus-Mesh with a very small diameter. This in turn reduced the number of hops a message takes thus lowering the total energy consumption.

\section{Conclusion \& Future Work}
The paper addresses vertex-centric graph processing on a near-memory oriented many-core system containing thousands of processing cores coupled with local memory. It introduces a programming model called the \textit{diffusive programming model} that allows spawning tasks from within the vertex data at runtime. These tasks are created in the form of \textit{actions} that contain the work that is sent in the form of messages to where the vertex resides. This enables graph algorithms to be implemented in asynchronous and message-driven way. The paper also presents an innovative vertex-centric data-structure that parallelizes both out and in-degree load of vertex objects across many cores and yet provides a single programming abstraction to the vertex objects.

Since the data structure is flexible and can grow and shrink a logical future direction is to design and implement dynamic graph algorithms. This can be implemented by having messages carrying \textit{actions} that mutate the graph structure. For example an \textit{action} containing new edges to be inserted in the graph or modifying or deleting existing edges. When the \textit{action} finishes modifying the graph structure it can invoke a computation, such as BFS, that re-compute from there without starting the execution all the way from scratch.

\begin{acks}
\textbf{To be added:}
\end{acks}

\bibliographystyle{ACM-Reference-Format}
\bibliography{Reference}


\appendix

\section{Research Methods}
The simulator that was designed to implement the work discussed in the paper is \texttt{CCASimulator}. It is available at
\begin{itemize}
    \item \href{https://github.com/bibrakc/CCA-Simulator}{https://github.com/bibrakc/CCA-Simulator}.
    \item Commit: /commit/479b4de916c54422a5dc36f687ead7598a2bf1a8
\end{itemize}

\subsection{How to Run}
The experiments can be reproduced by running the scripts provided in ``\texttt{/Papers/ICPP\_2024}'' directory. The ``\texttt{Runs}'' directory contains sub-directories for each dataset and application pair. The run script are provided with in these directories separately. Please insure you have set the correct path to the \texttt{repository} and \texttt{dataset} variables. The scripts will first compile the code and then execute that particular experiment based on the configuration. 

\textbf{Datasets:} The datasets used in the experiments will be provided in the evaluation phase. There are some example datasets in the \texttt{/Input\_Graphs} directory which you can use to get the setup working at your end. You can also go into the applications in the \texttt{Applications} directory and directly run the applications using the provided example input graphs in the \texttt{/Input\_Graphs}.

\subsection{Plotting and Analysing}
Once the experiments are done there will be separate file generated for each configuration and trial for that configuration. There is some randomness involved in the execution therefore we perform a number of trials and use the minimum for that configuration. 
There is a script that consolidates all the key parameters from the simulation into a single csv file for that particular dataset and application runs. It is \texttt{/Scripts/read\_all\_consolidate\_results.py}. Make sure to set the path in the script. It will read all the log files and generate a single csv file. For your convenience and reproducibility the csv file used for the paper is provided for each dataset and application pair in their respective directories. All the log files will be provided in the evaluation phase. These are about 380 MB zipped. Finally, we query these csv files using the \texttt{query\_csv\_*.py} scripts provided in \texttt{/Papers/ICPP\_2024/Scripts/}.

Plotting scripts along with the actual numbers used in the experiments are also provided. These are in the \texttt{/Papers/ICPP\_2024/Plots} directory.

\section{Online Resources}
\href{https://github.com/bibrakc/CCA-Simulator}{https://github.com/bibrakc/CCA-Simulator}

\end{document}